\documentclass{llncs}
\usepackage[utf8]{inputenc}
\usepackage{graphicx}
\usepackage{caption}
\begin{document} 

\title{The Bourgeois Gentleman, Engineering and Formal Methods}
\titlerunning{Formal Method in Industry}  
%
\author{Thierry Lecomte\inst{1}}
\authorrunning{Thierry Lecomte} 
%
\tocauthor{Thierry Lecomte}
\institute{ClearSy, 320 avenue Archimède, Aix en Provence, France\\
\email{thierry.lecomte@clearsy.com}}

\maketitle              

\begin{abstract}
Industrial applications involving formal methods are still exceptions to the general rule. Lack of understanding, employees without proper education, difficulty to integrate existing development cycles, no explicit requirement from the market, etc. are explanations often heard for not being more formal. 
This article reports some experience about a game changer that is going to seamlessly integrate formal methods into safety critical systems engineering.

\keywords{B method, safety platform, automated proof}
\end{abstract}
\section{Introduction}
The Moliere's Bourgeois Gentleman claimed that "for more than forty years [he has] been speaking prose while knowing nothing of it". What about imagining engineers claiming the same assertion but about formal methods ? Formal methods and industry are not so often associated in the same sentence as the formers are not seen as an enabling technology but rather as difficult to apply and linked with increased costs. Lack of understanding, employees without proper education, difficulty to integrate existing development cycles, no explicit requirement from the market, etc. are explanations often heard for not being more formal.
Our experience with formal methods, accumulated over the last 20 years \cite{DBLP:journals/entcs/Benveniste11}\cite{DBLP:journals/corr/BurdyDP17}\cite{DBLP:conf/fm/Lecomte08}\cite{DBLP:conf/fmics/Lecomte09}\cite{DBLP:journals/corr/abs-1210-6815}\cite{DBLP:conf/rssrail/Sabatier16}, clearly indicates that not every one is able to abstract, refine, and prove mathematically. The Swiss psychologist Piaget claimed that only one third of the population is able to handle abstraction\footnote{Skill acquired and developed during the so-called Formal Operational Stage}. However we are firmly convinced that formal methods are a fundamental key to ensure safety for our all-connected world.    
Several years ago, we imagined a new solution smartly combining the B formal method, a diverse compilation tool-chain and a double processor architecture \cite{lecomte2016double}. At that time, our sole objective was to reduce development costs but since then, given the full automation of the process, we are now considering this it as a way to obtain quite easily control-command systems certifiable at the highest levels of safety.
This paper briefly presents the technical principles of this platform, the successful experiments/deployments/dissemination before listing the remaining scientific and technological challenges to address in the future.
\section{CLEARSY Safety Platform}
The CLEARSY Safety Platform (abbreviated as CSSP in the rest of the document) is a new technology, both hardware and software, combining a software development environment based on the B language and a secured execution hardware platform, to ease the development of safety critical applications. \\
It relies on a software factory that automatically transforms  function into binary code that runs on redundant hardware. The starting point is a text-based, B formal model that specifies the function to implement. This model may contain static and dynamic properties that define the functional boundaries of the target software.  The B project is automatically generated, based on the inputs/outputs configuration (numbers, names). From the developer point of view, only one function (name \textit{user\_logic}) has to be specified and implemented properly. 
The implementable model is then translated using two different chains:
\begin{itemize}
\item Translation into C ANSI code, with the C4B Atelier B code generator (instance I$_1$). This C code is then compiled into HEX\footnote{a file format that conveys binary information in ASCII text form. It is commonly used for programming micro-controllers} binary code with an off-the-shelf compiler (gcc).
\item Translation into MIPS Assembly then to HEX binary code, with a specific compiler developed for this purpose (instance I$_2$). The translation in two steps allows to better debug the translation process as a MIPS assembly instruction corresponds to a HEX line.
\end{itemize}

\begin{center}
\includegraphics[width=\textwidth]{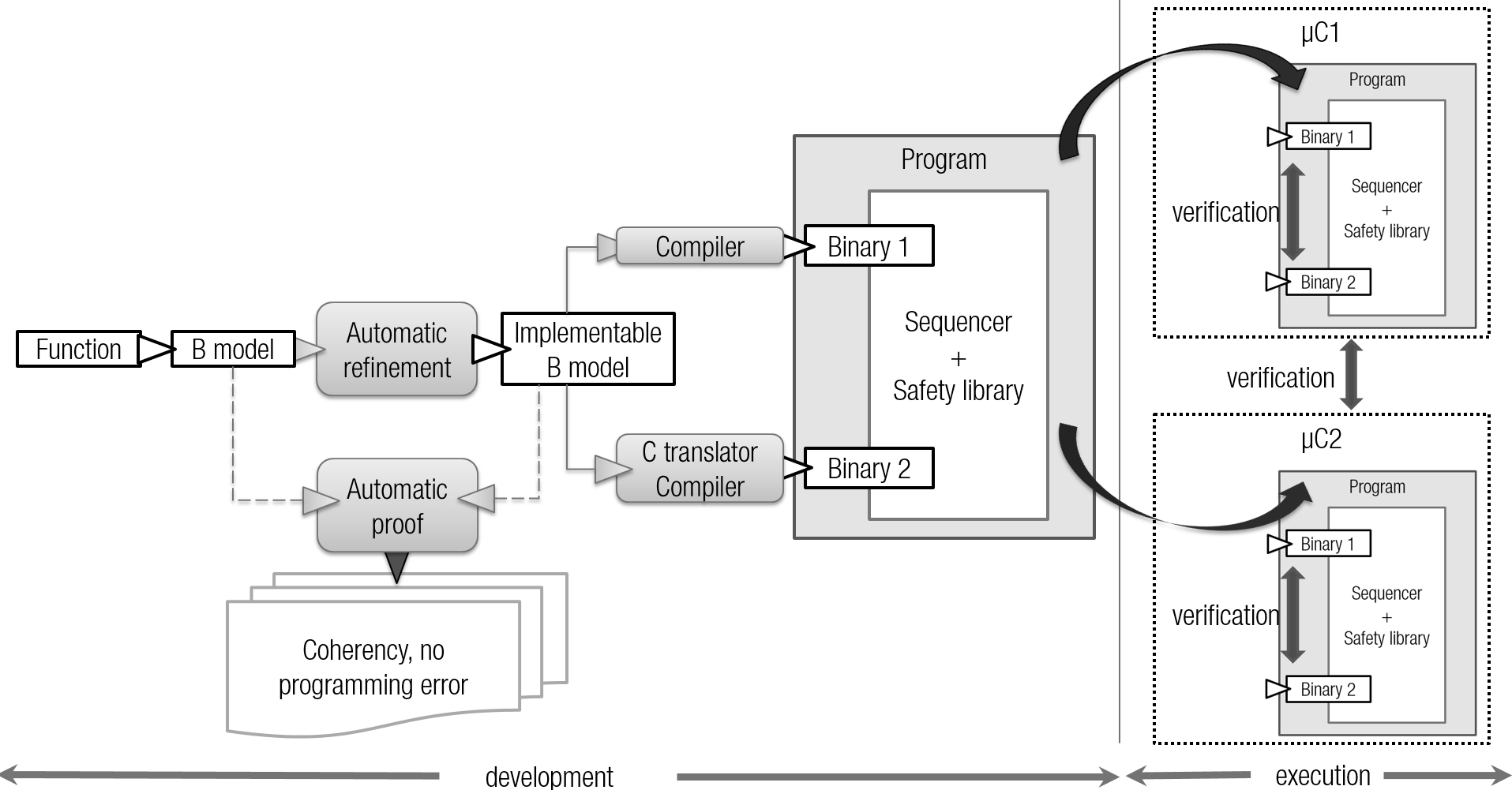}
\captionof{figure}{The safe generation and execution of a function on the double processor.}
\label{LCHIP-principe}
\end{center}
The software obtained is the uploaded on the execution platform to be executed by two micro-controllers.

\subsection{Safety}
These two different instances I$_1$ and I$_2$ of the same function are then executed in sequence, one after the other, on two PIC32 micro-controllers. Each micro-controller hosts both I$_1$ and I$_2$, so at any time 4 instances of the function are being executed on the micro-controllers. The results obtained by I$_1$ and I$_2$ are first compared locally on each micro-controller then they are compared between micro-controllers by using messages. In case of a divergent behaviour (at least one of the four instances exhibits a different behaviour), the faulty micro-controller reboots. 

The sequencer and the safety functions are developed once for all in B by the IDE design team and come along as a library. This way, the safety functions are out of reach of the developers and cannot be altered.
\begin{center}
\includegraphics[width=\textwidth]{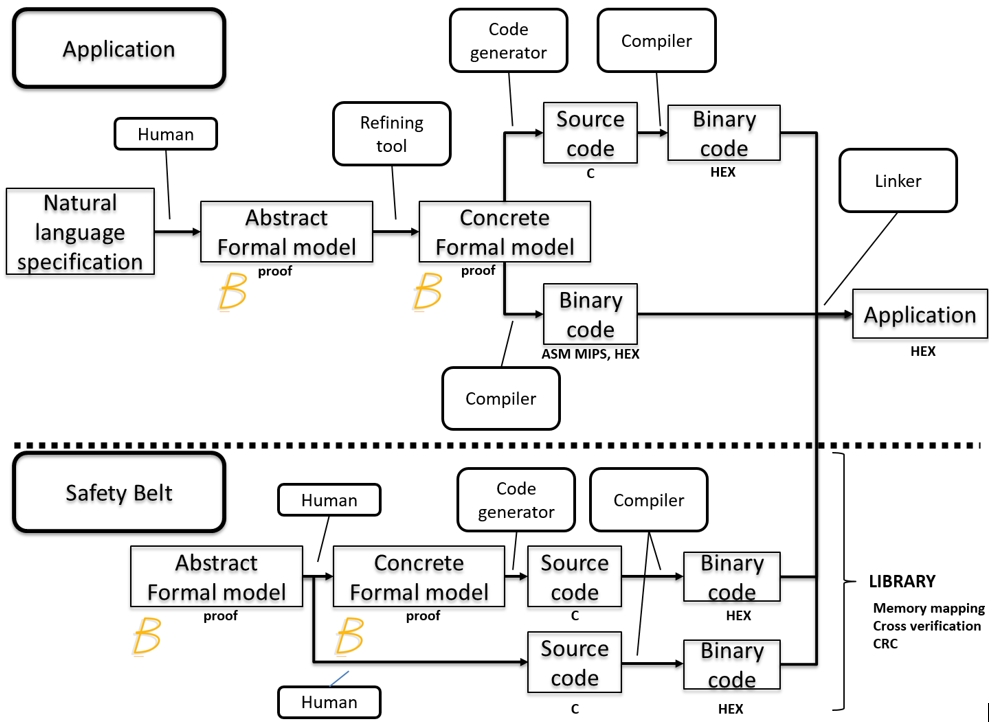}
\captionof{figure}{Process for developing an application and the safety library. Both application and safety belt rely on the B method plus some handwritten code - mainly I/O.}
\label{LCHIP-process}
\end{center}
The safety is based on several features such as:
\begin{itemize}
    \item the detection of a divergent behaviour,
    \item micro-controller liveness regularly checked by messages,
    \item the detection of the inability for a processor to execute an instruction properly\footnote{all instructions are tested regularly against an oracle},
    \item the ability to command outputs\footnote{outputs are read to check if commands are effective, a system not able to change the state of its outputs has to shutdown},
\end{itemize}

\begin{center}
\includegraphics[width=\textwidth]{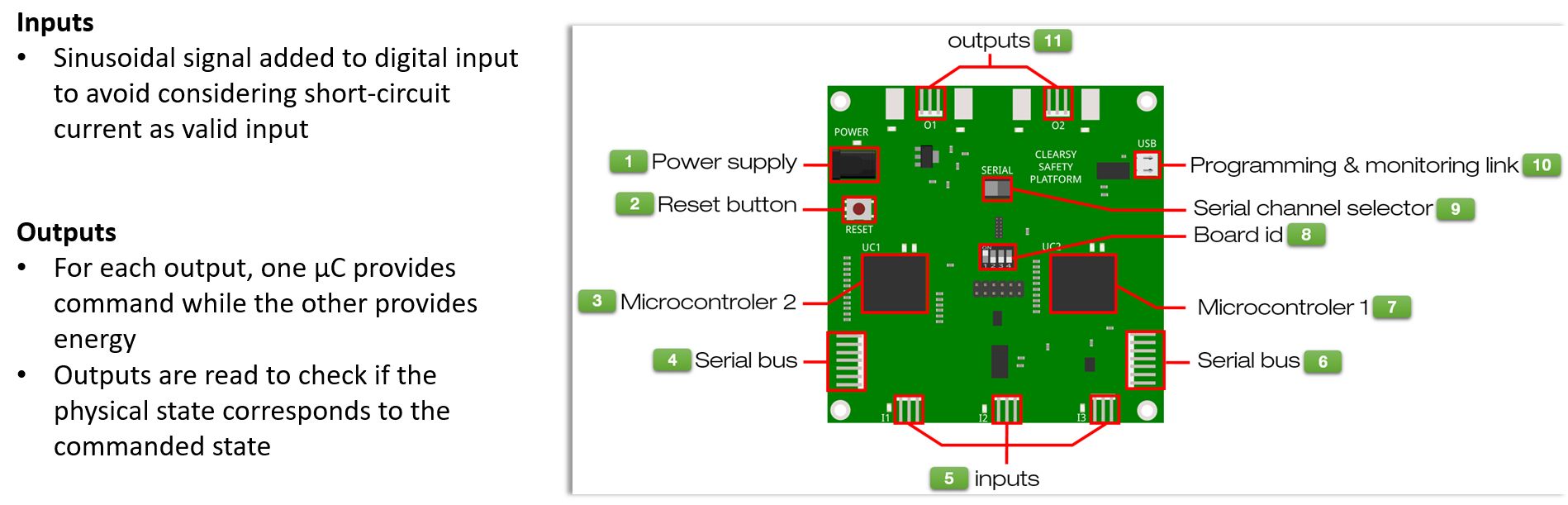}
\captionof{figure}{Some safety principles added to the hardware interface.}
\label{hw-safety-principles}
\end{center}

\begin{itemize}
    \item memory areas (code, data for the two instances) are also checked (no overlap, no address outside memory range),
    \item each output needs the two micro-controllers to be alive and providing respectively power and command, to be active (permissive mode). In case of misbehaviour, the detecting micro-controller deactivate its outputs and enter an infinite loop doing nothing.
\end{itemize}

\begin{center}
\includegraphics[width=\textwidth]{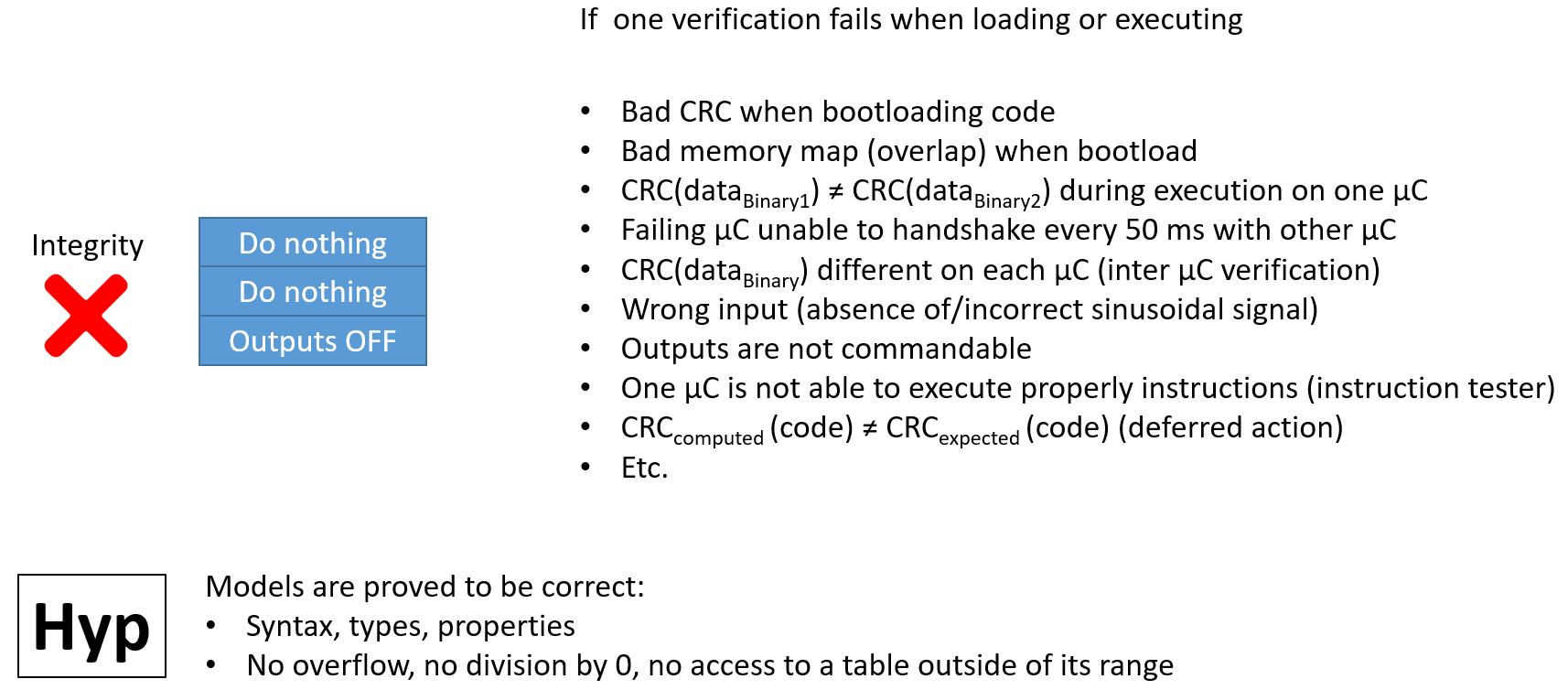}
\captionof{figure}{The verifications performed by the CLEARSY Safety Platform.}
\label{safety-principles}
\end{center}

\subsection{Target applications}
The execution platform is based on two PIC32 micro-controllers\footnote{PIC32MX795F512L providing 105 DMIPS at 80MHz}. The processing power available is sufficient to update 50k interlocking Boolean equations per second, compatible with light-rail signalling requirements. The execution platform can be redesigned seamlessly for any kind of mono-core processor if a higher level of performance is required. \\
The IDE provides a restricted modelling framework for software where:
\begin{itemize}
\item No operating system is used.
\item Software behaviour is cyclic (no parallelism).
\item No interruption modifies the software state variables.
\item Supported types are Boolean and integer types (and arrays of).
\item Only bounded-complexity algorithms are supported (the price to pay to keep the proof process automatic).
\end{itemize}

\section{Deployment and dissemination}

\subsection{Research and development}
CSSP was initially an in-house development project before being funded\footnote{The project is partly funded by BPI France, Région PACA, and Métropole Aix-Marseille, with a strong support from the “Pôles de compétitivité” I-Trans (Lille), SCS (Aix en Provence) and Systematic (Paris).} by the R\&D project LCHIP\footnote{Low Cost High Integrity Platform.} to obtain a generic version of the platform.
This project is aimed at allowing any engineer to develop a function by using its usual Domain Specific Language and to obtain this function running safely on a hardware platform. With the automatic development process, the B formal method will remain "behind the curtain" in order to avoid expert transactions. As the safety demonstration does not require any specific feature for the input B model, it could be handwritten or the by-product of a translation process. 

\begin{center}
\includegraphics[width=\textwidth]{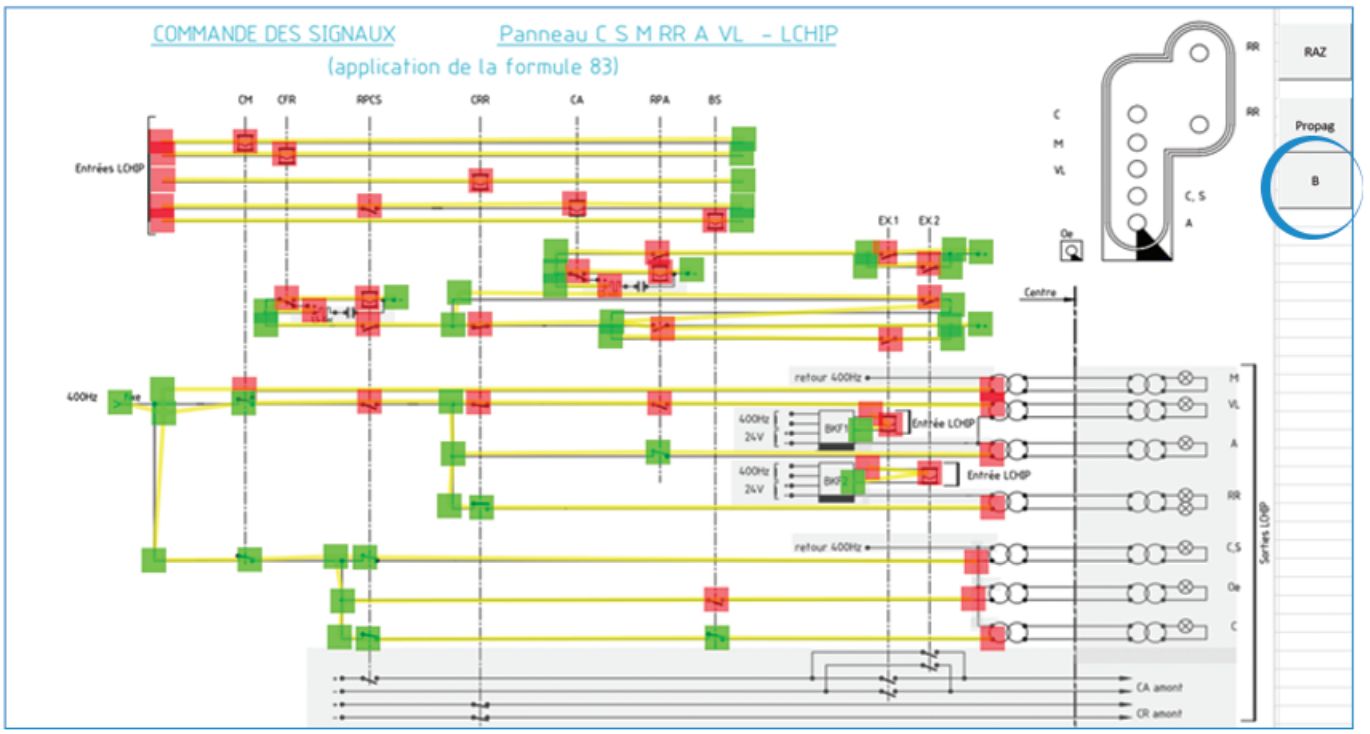}
\captionof{figure}{On-the-
y automatic model extraction from relay-based schematics and B model generation.}
\label{dsl}
\end{center}

Several DSL are planned to be supported at once (relays schematic, grafcet) based on an Open API  (Bxml). The translation from relays schematic is being studied for the French Railways.
The whole process, starting from the B model and finishing with the software running on the hardware platform, is expected to be fully automatic\footnote{"Push button" - the Graal of industry.}, even with "not so simple models" with the integration of the results obtained from some R\&D projects\footnote{to improve automatic proof performances (ANR-BWARE)}. 

\subsection{Education}

The IDE is based on Atelier B 4.5.3, providing a simplified process-oriented GUI. 

\begin{center}
\includegraphics[width=\textwidth]{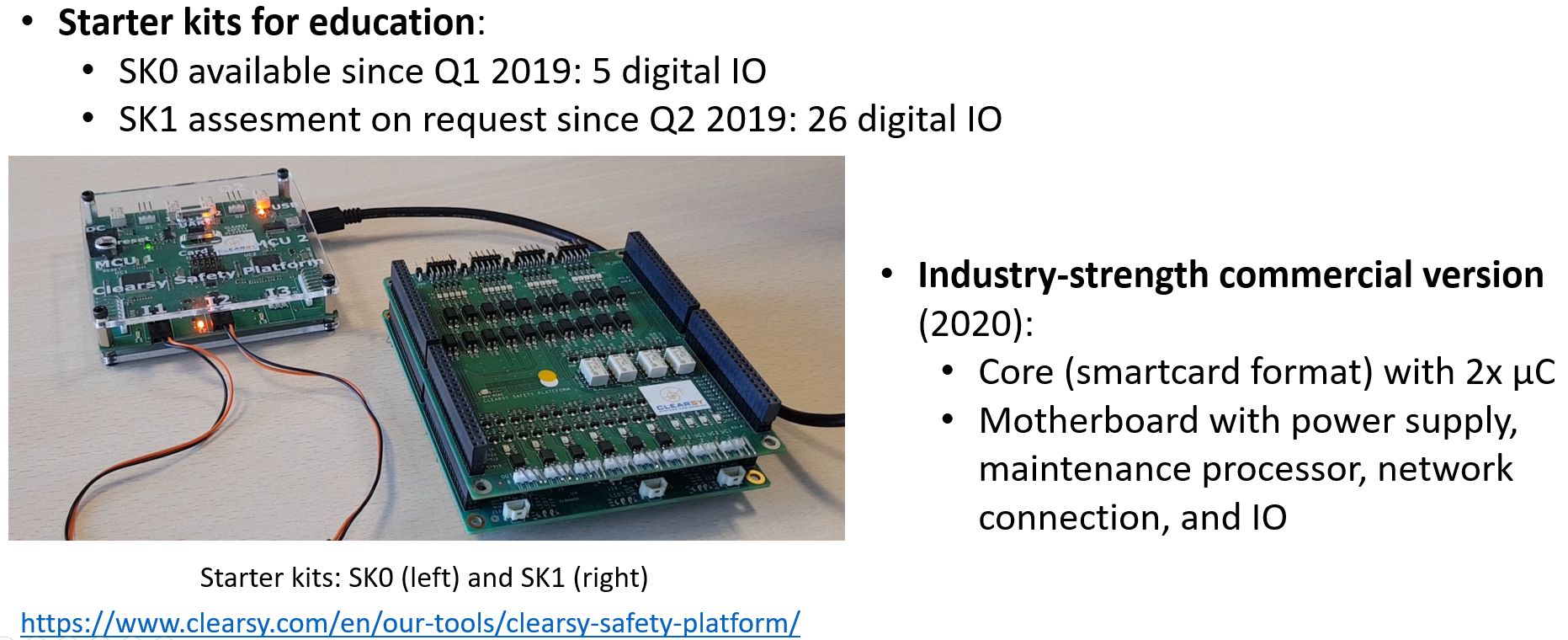}
\captionof{figure}{Starter kits for education.}
\label{sdk}
\end{center}

A first starter kit, SK$_0$, containing the IDE and the execution platform, was released by the end of 2017\footnote{https://www.clearsy.com/en/our-tools/clearsy-safety-platform/}, presented and experimented at the occasion of several hands-on sessions organised at university sites in Europe, North and South America. Audience was diverse, ranging from automation to embedded systems, mecatronics, computer science and formal methods. Results obtained are very encouraging : 
\begin{itemize}
    \item teaching formal methods is eased as students are able to see their model running in and interacting with the physical world, 
    \item less theoretic profiles may be introduced/educated to more abstract aspects of computation,
    \item the platform has demonstrated a certain robustness during all these manipulations and has been enriched with the feedback collected so far.
    \item CSSP is yet used to teach in M2 in universities and engineering schools.
\end{itemize}

\begin{center}
\includegraphics[width=\textwidth]{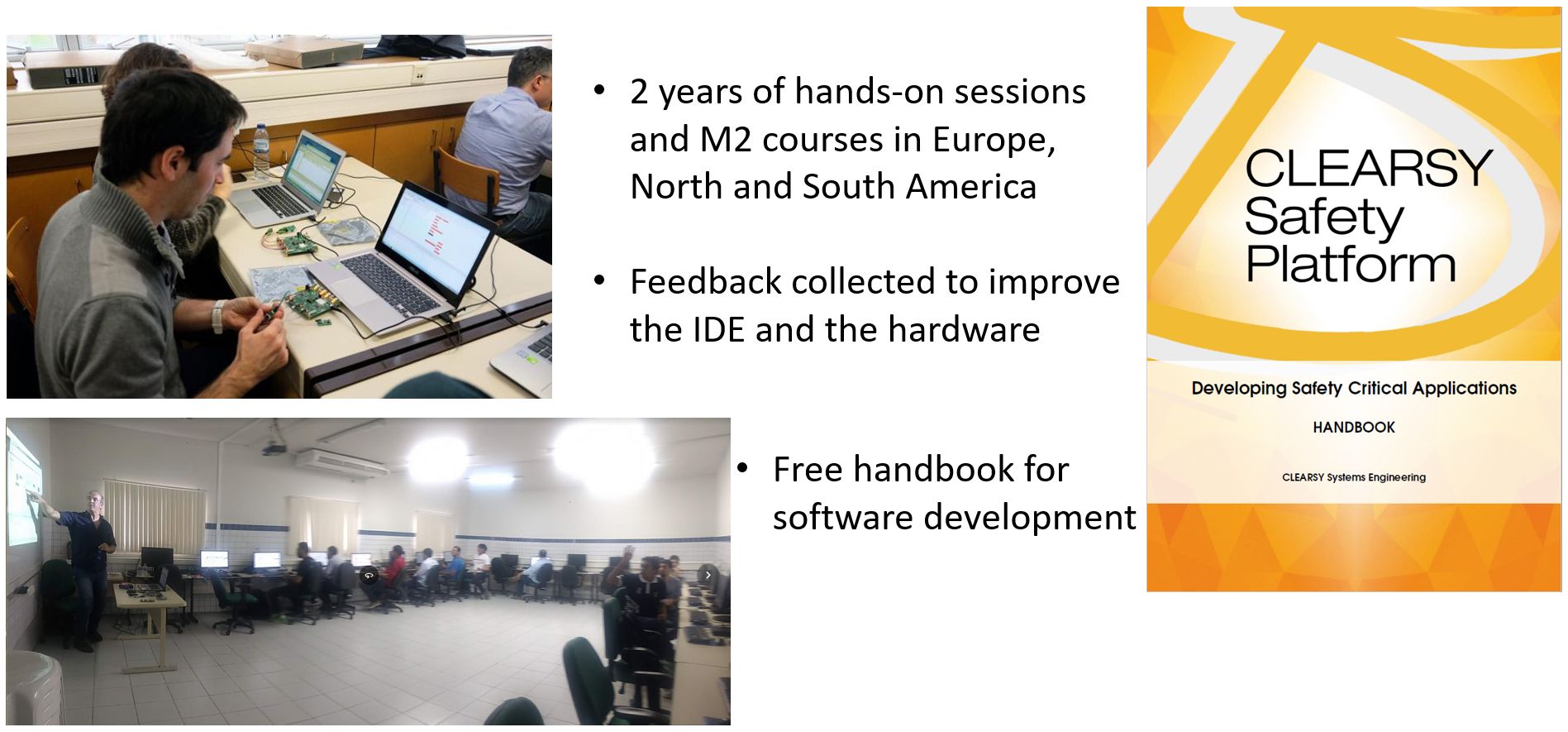}
\captionof{figure}{Exploitation and dissemination.}
\label{dissemination}
\end{center}

\subsection{Deployment}

CSSP building blocks are operating platform-screen doors in São Paulo L15 metro (certified in 2018 at level SIL3 by CERTIFER on the inopportune opening failure of the doors), in Stockholm City line (certified in 2019), and in New York city (to be certified in 2019). \\
A new starter kit, SK$_1$, released end of 2018 and aimed at prototyping\footnote{It embeds 20 inputs and 8 outputs, all digital.}, has been experimented by the French Railways for transforming  wired logic into programmed ones \cite{inbook} (track side signal control, wrong way temporary signalling system). This starter kit definitely attracts a lot of attention from industry, from railways but also robotics and autonomous vehicles.  
With the forthcoming CSSP Core (safety programmable logic controller) by the end of 2019, more deployments in industry are expected.

\section{Conclusion and Perspectives}
CSSP, combined with improved proof performances and connection with Domain Specific Languages, pave the way to easier development of SIL4 functions (including both hardware and software). The platform safety being out of reach of the software developer, the automation of the redundant binary code generation process and the certificates already obtained for products embedding CSSP building blocks, would enable the repetition of similar performances without requiring highly qualified engineers.\\
Moreover, the hardware platform is generic enough to host a large number of complexity-bounded\footnote{target complexity is lower than a metro automatic pilot one's} 
industry applications, with a special focus on the robotics and autonomous vehicles/systems  domains. \\
However some aspects have to be considered in the near future to ensure a wide dissemination in the target application domains like: 
\begin{itemize}
    \item improved automatic proof performances to reach 100 \% for not-so-complicated software functions, 
    \item support for continuous values (as opposed to digital ones), 
    \item support of more powerful, single-core processors, \item increase of the genericity while keeping the ability to be certified, 
    \item etc. 
\end{itemize}

\bibliographystyle{splncs03}
\bibliography{biblio}

\end{document}